\documentclass[14pt]{article}
\usepackage{amsmath}
\usepackage{amsfonts}
\usepackage{epsfig}

\def\Iden{\mbox{$\bf 1\ $}}

\begin{document}

\baselineskip .7cm

\title{Coherent states in Magnetic Resonance}
\author{Navin Khaneja \thanks{To whom correspondence may be addressed. Email:navinkhaneja@gmail.com} \thanks{Systems and Control Engineering, IIT Bombay.}}

\maketitle
\vspace{.1 in}

\begin{abstract}
{\rm In NMR experiments, interaction of quantized radio-frequency (rf) field leads to entanglement of nuclear spin with the electromagnetic field. 
In an entangled state, the nuclear spins are depolarized with no net magnetization, which cannot give a detectable signal in inductive detection. 
We show that when the electromagnetic field is in coherent state, inductive detection is just true. 
We develop the mathematics to study the evolution of a coherent rf-field with a 
sample of all polarized spins. We show that evolution can be solved in closed form as a separable state
of rf-field and spin ensemble, where spin ensemble evolves according to Bloch equations in an rf field.
We extend the analysis and results to a spin ensemble with Boltzmann polarization. The rabi frequency and coupling strength of spins to rf-field
depends on number state of the rf-field. We show that in interaction with a coherent rf-field, this variation in coupling strength, introduces negligible error.}
\end{abstract}

\section{Introduction}

Consider the Jaynes-Cummings Hamiltonian of a spin $\frac{1}{2}$ with coherent rf- field.

The Hamiltonian takes the form

$$ H = \omega a_n^{-}a_n^{+}  +  \kappa (a_n^{-}b_n^{+} + a_n^{+}b_n^{-})  + \omega S_z. $$

We index the number states of the rf-field as $| n \rangle$ such that $a_n^{+} |n \rangle = \sqrt{n+1} | n+1 \rangle$,
and $a_n^{-} |n \rangle = \sqrt{n} |n-1 \rangle$, where $a_n^{+}$ and $a_n^{-}$ are creation and annihilation operators for the field and $b^{+}$ and $b^{-}$ are creation and annihilation operators for the spin \cite{Loudon, Scully, Cohen}.

$$ a_n^{-}a_n^{+} |n \rangle = n+1 \ |n \rangle. $$

The spin up and down states are labelled as $|0 \rangle$ and $|1 \rangle$ with $S_z |0 \rangle = -\frac{1}{2}$, and $S_z | 1 \rangle = \frac{1}{2}$.

Consider the evolution of the state $a_{n-1}(t) |n-1 \rangle |1 \rangle + a_{n}(t) |n \rangle |0 \rangle $. This gives

\begin{eqnarray}
\label{eq:aeq}
\frac{d}{dt} \left [ \begin{array}{c} a_{n-1} \\ a_{n} \end{array} \right ] = -i \left [ \begin{array}{cc} (n + \frac{1}{2}) \omega & \kappa \sqrt{n} \\ \kappa \sqrt{n} & (n + \frac{1}{2}) \omega \end{array} \right ]\left [ \begin{array}{c} a_{n-1} \\ a_{n} \end{array} \right ].
\end{eqnarray}

The evolution of the state can be calculated. Starting from initial state $|n \rangle | 0 \rangle $, evolution for
time $t = \frac{\pi}{4 \kappa \sqrt{n}}$, where $n$, is photon number
generates a state (we neglect global phase),

\begin{equation}
\label{eq:ev1}
| \psi \rangle  = \frac{1}{\sqrt{2}} ( |n \rangle | 0 \rangle -
i |n-1 \rangle | 1 \rangle ).
\end{equation}

The corresponding density matrix is

$$ \rho_{fs} = | \psi \rangle \langle  \psi | = \frac{1}{2} (| n \rangle | 0 \rangle \langle 0 | \langle n | + | n-1 \rangle | 1 \rangle \langle 1 | \langle n-1 | -i | n \rangle | 0 \rangle \langle 1 | \langle n-1 | + i| n-1 \rangle | 1 \rangle \langle 0 | \langle n | ).$$

We take partial trace with respect to light field. Giving us a density matrix that takes the form $ \rho_s = \frac{1}{2} \Iden $. Taking expectation, gives $\langle S_x \rangle =0 $ and $\langle S_y \rangle = 0$, with gives undetectable magnetization.

We show that we can rectify the situation, if we assume the state of the light field is a coherent state.

The state of the light field and spin is $ | \psi \rangle |0 \rangle $, where $ | \psi \rangle$ is a {\it coherent state},

$$ | \psi \rangle = \sum_{n=0}^{\infty} \frac{\alpha^{\frac{n}{2}}}{\sqrt{n!}} \exp(-\frac{|\alpha|}{2})|n \rangle, $$  with $|\alpha| = \langle n \rangle$, the mean photon number.

In Eq. (\ref{eq:aeq}), starting from initial state $|n \rangle | 0 \rangle $, evolution for
time $t = \frac{\pi}{4 \kappa \sqrt{\langle n \rangle}}$, 

$$ |n \rangle | 0 \rangle  \rightarrow \cos (\frac{\pi n}{4 \langle n \rangle}) |n \rangle |0 \rangle -i \sin (\frac{\pi n}{4 \langle n \rangle}) |n-1 \rangle |1 \rangle, $$

which for $n$ around $\langle n \rangle$ reduces to Eq. $\ref{eq:ev1}$.

The  photon-spin state $ | \psi \rangle |0 \rangle $ evolves to

$$ \frac{| \psi_A \rangle |0 \rangle -i | {\psi}_B \rangle \exp(-i \omega) | 1 \rangle}{\sqrt 2}, $$

where,

$$  | \psi_A \rangle = \sqrt{2} \sum_{n=1}^{\infty}  \cos (\frac{\pi n}{4 \langle n \rangle}) \exp(-i (n + \frac{1}{2})\omega) \frac{\alpha^{\frac{n}{2}}}{\sqrt{n!}} |n \rangle, $$

$$ | \psi_B \rangle = \sqrt{2} \sum_{n=1}^{\infty}  \sin (\frac{\pi n}{4 \langle n \rangle}) \exp(-i (n - \frac{1}{2})\omega) \frac{\alpha^{\frac{n}{2}}}{\sqrt{n!}} |n-1 \rangle. $$

when probability in $| \psi \rangle$ is concentrated around $|\alpha| = \langle n \rangle$, $| \psi_A \rangle $ and $| {\psi}_B \rangle $ are collinear, we can write the above state as

$$ | \psi_A \rangle ( \frac{|0 \rangle -i \exp(-i \omega) |1 \rangle}{\sqrt{2}}). $$

Partial trace wrt to field states gives $ \rho_s = \frac{(|0 \rangle -i  \exp(-i \omega) |1 \rangle )( \langle 0 | - i
\exp(i \omega) | \langle 1 | )}{2} $, which is just transverse magnetization.

The product
$$ \langle {\psi}_B | \psi_A \rangle = \sum_{n=1}^{\infty} 2 \cos (\frac{\pi n}{4 \langle n \rangle}) \sin (\frac{\pi (n+1)}{4 \langle n \rangle}) \exp(-|\alpha|) \frac{\alpha^{\frac{1}{2}}}{\sqrt{n+1}}\frac{\alpha^{n}}{n!} = \sum_{n \in E}^{\infty} + \sum_{n \in E^C}^{\infty}, $$

where $$ E = [ \langle n \rangle \pm \langle n \rangle ^{\frac{3}{4}} ], $$

$$ \sum_{E} 2 \cos (\frac{\pi n}{4 \langle n \rangle}) \sin (\frac{\pi (n+1)}{4 \langle n \rangle}) \exp(-|\alpha|) \frac{\alpha^{\frac{1}{2}}}{\sqrt{n+1}}\frac{\alpha^{n}}{n!} \geq   (1 - \delta) \sum_{E} 2 \cos (\frac{\pi n}{4 \langle n \rangle}) \sin (\frac{\pi (n+1)}{4 \langle n \rangle}) \exp(-|\alpha|) \frac{\alpha^{n}}{n!}, $$

where $\sqrt{\frac{\alpha}{n+1}} > 1 - \delta$ for $n \in E$. Writing  $2 \cos (\frac{\pi n}{4 \langle n \rangle}) \sin (\frac{\pi (n+1)}{4 \langle n \rangle})$ as $\sin(\frac{\pi (n+\frac{1}{2})}{2 \langle n \rangle})  - \sin(\frac{\pi}{4 \langle n \rangle})$, we have

$$ \sum_E \sin(\frac{\pi (n+\frac{1}{2})}{2 \langle n \rangle})  - \sin(\frac{\pi}{4 \langle n \rangle}) \exp(-|\alpha|) \frac{\alpha^{n}}{n!}  = \sum_E 1 - \frac{\pi^2}{2^2}\underbrace{\frac{(n - \langle n \rangle + \frac{1}{2})^2}{\langle n \rangle ^2}}_{\leq \epsilon}  \exp(-|\alpha|) \frac{\alpha^{n}}{n!}  \geq (1 - \epsilon)(1 - P(E^c)), $$

$$ \cos \theta = | \langle {\psi}_B | \psi_A \rangle | \geq (1 - \epsilon')(1 - P(E^c)) - 2 P(E^c) = 1 - 3 P(E^c), $$

$ P(E^C) \leq \frac{1}{\sqrt{n}} $.

We again compute the partial trace of density matrix $\rho_{fs}$, corresponding to the state $ \frac{| \psi_A \rangle |0 \rangle -i | {\psi}_B \rangle \exp(-i \omega) | 1 \rangle}{\sqrt 2}$.

Abbreviating $A = \psi_A$ and $B = \psi_B$, let ${A}^{\perp}$ be unit vector perpendicular to $A$ in the $A-B$ plane. Then for $ \sin \theta = | \langle B | A^{\perp} \rangle |$,

$$ \rho_{fs} = \frac{| A \rangle |0 \rangle \langle 0 | \langle A |  + | B \rangle | 1 \rangle \langle 1 | \langle B | + i \exp(-i \omega) | A \rangle |0 \rangle \langle 1 | \langle B | -i \exp(i \omega) | B \rangle |1 \rangle \langle 0 | \langle A | }{2}. $$

Partial trace is 
$$ \rho = \frac{1}{2} \left [ \begin{array}{cc} 1 & \cos \theta \exp(j \phi) \\ \cos \theta \exp(-j \phi) & \cos^2 \theta + \sin^2 \theta \end{array} \right ], $$ where $\exp(j \phi) = i \exp(-i \omega)$. 

The magnetization $\langle S_x \rangle = tr (\rho S_x) = \cos \theta \cos \phi $
and $ \langle S_y \rangle = tr (\rho S_y) = \cos \theta \sin \phi$.
Thus we have shown that if we have a spin $\frac{1}{2}$ driven by a rf-field in coherent state, the spin state remains only very weakly entangled with the coherent state and we can do a detection of its transverse magnetization.

A quick recapitulation of partial trace and its significance. Let $\rho_{AB}$ be the density matrix of a bipartite quantum system. Let $| i \rangle$ and $|j \rangle $ be orthonormal basis for system $A$ and $B$. $| i \rangle \otimes |j \rangle $ are basis for the bipartite system $AB$. If the ensemble indexed by $k$ is made of pure states $\sum_{ij} a^k_{ij} |i \rangle |j \rangle$, the density matrix for the bipartite system takes the form

$$ \rho_{AB} = \sum_k p_k (\sum_{ijlm}a^k_{ij}a^{*k}_{lm} |i \rangle \langle l| \otimes | j \rangle \langle m |. $$
Local Measurement on subsystem $A$, by $M \otimes \Iden$ is given by

\begin{eqnarray}
\langle M \rangle &=& tr((M \otimes I)\rho_{AB}) \\
 &=& \sum_{il} \underbrace{\sum_k p_k \sum_{j=m} a^k_{ij}a^{*k}_{lm}}_{(\rho_A)_{il}} \langle l | M | i \rangle = tr(\rho_A M).
\end{eqnarray}

where $\rho_A$, is the partial trace obtained by tracing over system $B$, given
as

$$ (\rho_A)_{il} = \sum_j \langle i j | \rho_{AB} | l j \rangle. $$

\section{Evolution of all polarized spin ensemble in coherent rf-field}

We now consider a spin ensemble of $M$ spins, all polarized to begin with, interacting with a electromagnetic field in number state $| n \rangle$.
We choose a symmetrized basis for the spin states as
$$ |e_k \rangle = \frac{1}{\sqrt{^MC_{k-1}}}| 0^{M+1-k} 1^{k-1} \rangle, $$
where $| 0^{M+1-k} 1^{k-1} \rangle$ represents a state which represents all combinations, where $1$ appears in $k-1$ spots. For example
$|e_2 \rangle$ represents
$$ |e_2 \rangle = \frac{1}{\sqrt{M}} |1 0 \dots 0 \rangle + |0 1 0 \dots 0 \rangle + |0 \dots 0 1 \rangle. $$When a photon is absorbed it flips a spin. Starting from state $ |n\rangle |e_1 \rangle $, the general state of the photon and spin ensemble is written as

$$ \sum_{k=1}^{M+1} a_k |n - k + 1 \rangle |e_k \rangle. $$
We can write an equation for evolution of the coefficients $a_k$. The transition rate $a_{k, k+1}$ from $|n-k+1 \rangle |e_k \rangle$ to $|n-k \rangle |e_{k+1} \rangle$, ($a_{k,k+1} = a_{k+1, k}$), is

\begin{equation}
\label{eq:transitionrate}
\kappa \frac{\sqrt{^MC_{k}}}{\sqrt{^MC_{k-1}}} = \kappa \sqrt{n-k+1} \sqrt{k (M-k+1)} \sim
\kappa \sqrt{n} \sqrt{k (M-k+1)}.
\end{equation}
Then

\begin{equation}
  \label{eq:a.equation}
\dot{a} = -i \left [ \begin{array}{cccc} (n +1 - \frac{M}{2}) &  a_{1, 2} & \dots & a_{1, n} \\ a_{2, 1} & (n + 1 - \frac{M}{2}) & \hdots & \vdots\\ \hdots & \hdots & \ddots & \vdots \\
\hdots & \dots & a_{M-1, M} &  (n + 1 - \frac{M}{2}) \end{array} \right ]a.
\end{equation}
Let $b_k = \exp(i (k-1) \frac{\pi}{2})a_k$, then

\begin{equation}
  \label{eq:b.equation}
\dot{b} = \underbrace{\left [ \begin{array}{cccc} -i \omega(n + 1 - \frac{M}{2}) &  - a_{1, 2} & \dots & - a_{1, n} \\ a_{2, 1} & -i \omega(n + 1 - \frac{M}{2}) & \hdots & \vdots\\ \hdots & \hdots & \ddots & \vdots \\
\hdots & \dots & a_{M-1, M} &  -i \omega(n + 1 - \frac{M}{2}) \end{array} \right ]}_B b.
\end{equation}

We postulate a solution of the following form. $b_{k+1}(t)$ is the coefficient of $|e_{k+1} \rangle$ in

$$ \exp(-i t\omega (n + 1 - \frac{M}{2})) (\cos \kappa \sqrt n t |0 \rangle + \sin \kappa \sqrt n t |1 \rangle )^{\otimes M},   $$ where $^{\otimes M}$ denotes the $M$ fold tensor product.

The coefficient of $|e_{k+1} \rangle$ in

$$ (\cos \kappa \sqrt n t |0 \rangle + \sin \kappa \sqrt n t |1 \rangle )^{\otimes M},
$$

is $\sqrt{{M \choose k}} \cos^{M-k} \kappa \sqrt n t \sin^{k} \kappa \sqrt n t$.
Differentiating with $t$ gives $\dot{b}_{k+1}(t)=$

\begin{eqnarray*}
&&-i (n + 1 -\frac{M}{2}) b_{k+1}(t) - \sqrt{{M \choose k}} (M-k) \cos^{M-k-1}\kappa \sqrt n t \sin^{k+1} \kappa \sqrt n t  + \sqrt{{M \choose k}} k \cos^{M-k+1}\kappa \sqrt n t \sin^{k-1} \kappa \sqrt n t \\ &=& -\frac{\sqrt{{M \choose k}}M-k}{\sqrt{{M \choose (k+1)}}} b_{k+2}(t) + \frac{\sqrt{{M \choose k}} k}{\sqrt{{M \choose (k-1)}}} b_{k}(t) -i \omega (n + 1 - \frac{M}{2}) b_{k+1}(t) \\
&=& \kappa \sqrt{n} ( -\sqrt{(M-k)(k+1)} b_{k+2}(t) + \sqrt{(M-k+1)k}\ b_{k}(t)) + -i \omega (n + 1 - \frac{M}{2}) b_{k+1}(t).
\end{eqnarray*}

This gives $\bar a_{k+1}(t)$ as the coefficient of $|e_{k+1} \rangle$ in 
\begin{equation}
\label{eq:Aev}
(\cos \kappa \sqrt n t |0 \rangle - i \sin \kappa \sqrt n t |1 \rangle )^{\otimes M},
\end{equation} and $a_{k+1}(t) = \exp(-i \omega t (n + 1 - \frac{M}{2})) \bar a_{k+1}(t)$. 

If the initial state of the field is a coherent state, $$ | \psi \rangle = \sum_{n=0}^{\infty} \underbrace{\frac{\alpha^{\frac{n}{2}}}{\sqrt{n!}} \exp(-\frac{|\alpha|}{2})}_{c_n} |n \rangle, $$ with parameter $\alpha = |\alpha| e^{-j 2 \theta}$, then the spin-photon state evolves as

$$ \sum_k (\sum_{n} \exp(-i \omega t (n + k + 1 - \frac{M}{2})) c_{n+k} |n \rangle) \bar a_{k+1}(t)|e_{k+1} \rangle, $$ which can be written as 

\begin{equation}
  \label{eq:spinphoton.state}
  \sum_k \exp(-i \omega t (k + 1 - \frac{M}{2}) -i k \theta ) (\sum_{n} \exp(-i \omega t (n+\theta) ) \frac{|\alpha|^{\frac{n+k}{2}}}{\sqrt{n+k!}} \exp(-\frac{|\alpha|}{2}) | n \rangle) \bar a_{k+1}(t)|e_{k+1} \rangle.
  \end{equation}

For $n$ near mean photon number $|\alpha| = \langle n \rangle$, we have
$\frac{|\alpha|^{\frac{k}{2}}}{\sqrt{(n+k)\dots (n+1)}} \sim 1$. The statement can be formalized by taking the inner product between

$ | \psi_1 \rangle = \sum_{n} \exp(-i \omega t (n+\theta) ) \frac{|\alpha|^{\frac{n+k}{2}}}{\sqrt{n+k!}}  \exp(-\frac{|\alpha|}{2}) |n \rangle $ and 
$  | \psi_2 \rangle = \sum_{n} \exp(-i \omega t (n+\theta) ) \frac{|\alpha|^{\frac{n}{2}}}{\sqrt{n!}} \exp(-\frac{|\alpha|}{2}) |n \rangle, $ then

\begin{equation}
\label{eq:inpro}
\langle \psi_1 | \psi_2 \rangle = \sum_{n} \frac{|\alpha|^{\frac{k}{2}}}{\sqrt{(n+k)(n+k-1)..(n+1)}}  \frac{\alpha^{n}}{n!} \exp(-|\alpha|) = \sum_{E} + \sum_{E^c}.
\end{equation}
where $E = \{ |\alpha| - n', \dots, |\alpha| + n' \}$. On $E$, we have, $\frac{|\alpha|^{\frac{k}{2}}}{\sqrt{(n+k)(n+k-1)..(n+1)}} \geq \frac{(|\alpha|/\alpha')^{\frac{k}{2}}}{\sqrt{(1+ 1/\alpha')\dots(1 + k/\alpha')}}$, where say $\alpha' = |\alpha| + n'$.
Let $c_1 = \sqrt{(1+ 1/\alpha')\dots(1 + k/\alpha')}$, where $\ln c_1 = \frac{1}{2} \sum_j \ln(1 + \frac{j}{\alpha'})$, expanding
$\ln ( 1 + x) = x - \frac{x^2}{2} + \dots  < x$, which gives $\ln c_1 \leq \frac{k(k+1)}{2 \alpha'}$. 
Let $c_2 = (\alpha/\alpha')^{\frac{k}{2}}$, then $\ln c_2 = \frac{k}{2} \ln (1 - \frac{n'}{\alpha'}) \geq - \frac{k n'}{\alpha'}$, implying $c_2 \geq \exp(-k n'/\alpha')$.

Choosing, $n ' = \langle n \rangle ^{\frac{1}{2} + \beta}$, we get $\frac{c_2}{c_1} \geq \frac{\exp(-k n'/\alpha')}{\exp(\frac{k(k+1)}{2 \alpha'})}$. For $\ln \frac{k}{ \langle n \rangle} < \frac{1}{2}-\beta$, we get 
$\frac{c_2}{c_1} \geq 1 - \epsilon$. $Pr(E^c) < \langle n \rangle^{-2 \beta}$.

Therefore, for $\ln \frac{k}{ \langle n \rangle} < \frac{1}{2}-\beta$, we have,  $\langle \psi_1 | \psi_2 \rangle > 1 - \epsilon$.

Then the spin photon state in Eq. (\ref{eq:spinphoton.state}) can be written as

\begin{equation}
\label{eq:Cev}  
| \psi(t) \rangle \ (\cos \kappa \sqrt{\langle n \rangle} t |0 \rangle - i \exp(-i (\omega t + \theta) ) \sin \kappa \sqrt{\langle n \rangle} t |1 \rangle )^{\otimes M}.
\end{equation}

where  $| \psi(t) \rangle$ is a coherent state with parameter $ \alpha = |\alpha| \exp(i 2 \omega t).$ When evolution time is chosen such that $t = \frac{\pi}{4 \kappa \sqrt{\langle n \rangle}}$, the state takes the form

$$  | \psi(t) \rangle (\frac{|0 \rangle - i \exp(-i ( \omega t + \theta))|1 \rangle}{\sqrt 2} )^{\otimes M}. $$

The state of field and spin ensemble is a separable state. We can evaluate the net transverse magnetization and it takes the form

$$ \langle S_x + i S_y \rangle = M \cos(\omega t + \theta + \frac{\pi}{2}) + i \sin(\omega t + \theta + \frac{\pi}{2}). $$

\subsection{Error due to variation in coupling strength to different number states}

In Eq. \ref{eq:transitionrate} we approximated $ \sqrt{n-k+1} \sim \sqrt{n}$. Going from Eq. \ref{eq:Aev} to \ref{eq:Cev}, we made an approximation, $\sqrt{n} \sim \sqrt{\langle n\rangle}$. This may introduce an error when we analyze evolution of spin ensemble and coherent state as in Eq. (\ref{eq:spinphoton.state}) and subsequent analysis. We can capture the error in the evolution.
Given the equation $\dot{\tilde x} = (A + \Delta) \tilde x $, let $\dot{x} = A x$, solution to unperturbed part of the evolution. The difference of the evolution
$y = \tilde{x} -x$, takes the form  $\dot{y} = (A + \Delta)y  + \Delta x, $ where
$y = \int_0^T \exp(A + \Delta)(T-\tau) \Delta x$, then
$$ |\int_0^T \exp(A + \Delta)(T-\tau) \Delta x|^2 \leq (\int_0^T | \exp(A + \Delta)(T-\tau) \Delta x|)^2 \leq (\int_0^T |\Delta x|)^2 \leq T \int_0^T |\Delta x|^2   \leq T^2 |\Delta x|_{max}^2 $$
where $\exp(A + \Delta)$ is an orthogonal matrix. Let $\epsilon_{n, k+1} = \tilde b_{k+1}^n - b_{k+1}^n$
where $ \tilde b_{k+1}^n$ is the true evolution in Eq. (\ref{eq:b.equation}) and  $b_{k+1}^n$ is the evolution when we approximate $\sqrt{n} \sim \sqrt{\langle n\rangle}$.

$\dot{b}_{k+1}(t)=$

\begin{eqnarray*}
&=& \kappa \sqrt{n}(-\sqrt{(M-k)(k+1)} b_{k+2}(t) + \sqrt{(M-k+1)k}\ b_{k}(t) -i \omega (n + 1 - \frac{M}{2}) b_{k+1}(t)). 
\end{eqnarray*}

The approximation error can be bounded as

\begin{eqnarray*}
\kappa^2 (\sqrt{n-k+1} - \sqrt{|\alpha|})^2 &\leq& 2 \kappa^2 (\sqrt{n-k+1} - \sqrt n)^2 + 2 \kappa^2 (\sqrt{n} - \sqrt{|\alpha|})^2 \\
&\leq& \frac{2 \kappa^2 (k-1)^2}{n} + 2 \kappa^2 (\sqrt{n} - \sqrt{|\alpha|})^2.
\end{eqnarray*}

Using, $\sqrt{(M-k)(k+1)}b_{k+2} \sqrt{(M-k+1)k} b_{k} = b_{k+1}^2 k(M-k)$, we get 

\begin{eqnarray*}
(|\Delta b|_{k+1}^n)^2 &\leq& 2 \kappa^2 ( \frac{1
}{n} ( (k-1)^2 (M-k)(k+1)b_{k+2}^2 + (k-1)^2(M-k+1)k b_{k}^2 -2 (k-1)^2 b_{k+1}^2 k(M-k)) \\ 
&+& (\sqrt{n} - \sqrt{|\alpha|})^2)( (M-k)(k+1)b_{k+2}^2 + (M-k+1)k b_{k}^2 - 2 b_{k+1}^2 k(M-k))).
\end{eqnarray*}

Let $\epsilon_n^2 = \sum_{k} \epsilon_{n, k}^2$,

The coefficient $b_{k+1}^2$ is ${M \choose k} \underbrace{\cos^{2(M-k)}\kappa \sqrt n t}_{p^{M-k}} \underbrace{\sin^{2k} \kappa \sqrt n t}_{q^{k}}$. Let $X = k+1$, then

\begin{eqnarray*}
\frac{1}{T^2} \epsilon_n^2 &\leq& \frac{2 \kappa^2}{n} ( E(X-2)^2(M+1-X)X + E (X)^2(X+1)(M-X) -2 E(X-1)^2 X (M-X) 
) \\ 
&+& (\sqrt{n} - \sqrt{|\alpha|})^2( E(M+1-X)X  + E (X+1)(M-X) -2 E(X (M-X))).
 \end{eqnarray*} 

\begin{eqnarray*}
\frac{1}{T^2} \epsilon_n^2 &\leq& \frac{2 \kappa^2}{n} E X(X+2)(M-X) \\
&+& 2 \kappa^2 |\alpha| \frac{(\sqrt{n} - \sqrt{|\alpha|})^2}{|\alpha|}( E(M+1-X)X  + E (X+1)(M-X) -2 E(X (M-X)) ).
 \end{eqnarray*} 

\begin{eqnarray*}
\frac{1}{T^2} \epsilon_n^2 &\leq& \frac{2 \kappa^2}{n} M^3 \\
&+& 2 \kappa^2 (\sqrt{n} - \sqrt{|\alpha|})^2 M.
 \end{eqnarray*}

Calculating $E(\sqrt{n} - \sqrt{|\alpha|})^2$, over $n$, for $n \geq |\alpha|$, we have 

$$\sqrt{n} - \sqrt{|\alpha|} = \sqrt {|\alpha| + \delta n} - \sqrt{|\alpha|} = \sqrt{|\alpha|} (\sqrt {1 + \frac{\delta n}{|\alpha|}} - 1 ) \leq \frac{\delta n}{2 \sqrt{|\alpha|}}. $$
for $n \leq |\alpha|$, we have,

$$ \sqrt{|\alpha|} - \sqrt{n}  =  \sqrt{|\alpha|} - \sqrt {|\alpha| - \delta n} = \sqrt{|\alpha|} (1 - \sqrt{1 - \frac{\delta n}{|\alpha|}}) \leq \frac{\delta n}{2 \sqrt{|\alpha|}}. $$

Therefore, $E(\sqrt{n} - \sqrt{|\alpha|})^2 = E(\frac{\delta n^2}{|\alpha|}) = 1$.

Then writing the state in Eq. (\ref{eq:spinphoton.state}) with an error part $|\phi \rangle$ due to approximation, $\sqrt{n} \sim \sqrt{\langle n\rangle}$, we get 

$$ \sum_k (\sum_{n} \exp(-i \omega t (n + k + 1 - \frac{M}{2})) c_{n+k} |n \rangle) \bar a_{k+1}(t) |e_{k+1} \rangle + \underbrace{\sum_k (\sum_n \exp(-i \frac{\pi}{2}(k-1)) c_{n+k} \epsilon_{n+k, k} |n \rangle |e_{k+1} \rangle}_{|\phi \rangle}. $$

\begin{equation}
\label{eq:err1}
| | \phi \rangle |^2 = \sum_n c_n^2 \epsilon_{n}^2  = \sum_{n \geq |\alpha| - m \sqrt{|\alpha|} }c_{n}^2 \epsilon_{n}^2 + \sum_{0}^{|\alpha| - m \sqrt{|\alpha|}} c_{n}^2 \epsilon_{n}^2
\leq \sum_{n \geq |\alpha| - m \sqrt{|\alpha|} }c_{n}^2 \epsilon_{n}^2 + 4 \sum_{0}^{|\alpha| - m \sqrt{|\alpha|}} c_{n}^2,
\end{equation}where factor of $4$ in above Eq. (\ref{eq:err1}) comes about from fact that distance between two unit vectors is bounded by $2$.

We consider the sum $ s = \sum_{k = 0}^{|\alpha| - m \sqrt{|\alpha|}} \exp(-|\alpha|) \frac{|\alpha|^k}{k !} $, using stirling's approximation,

$ \frac{1}{n!} \leq \frac{1}{\sqrt{2 \pi}} \frac{e^{n}}{n^{n+\frac{1}{2}}}$, we have,

$$ s \leq \sum_{k = 0}^{|\alpha| - m \sqrt{|\alpha|}} \exp(-|\alpha|) \frac{|\alpha|^k e^k }{k^k}, $$ By using $\frac{|\alpha|^k e^k }{k^k}$, is an increasing function for $k \leq |\alpha|$, we obtain

$s \leq n \frac{|\alpha|^n e^n }{n^n}$, where $n =  {|\alpha| - m \sqrt{|\alpha|}}$. Let $c = (n/ |\alpha|)^n = (1 - \frac{m}{\sqrt{|\alpha|}})^n$,
$\ln c = n \ln (1 - \frac{m}{\sqrt{|\alpha|}}) \geq n (-\frac{m}{\sqrt {|\alpha|}} -\frac{2 m^2}{3 |\alpha|^2}) $, substituting we get
$s \leq (|\alpha| - m \sqrt{|\alpha|}) \exp^{\frac{-m^2}{3}}$. 

Let $T = \frac{\pi}{4 \kappa \sqrt{\langle n \rangle}}$, which is the time for a $\frac{\pi}{2}$ pulse.
The above error in Eq. \ref{eq:err1}, is bounded by $\frac{1}{|\alpha|}( \frac{M^3}{(|\alpha| - m \sqrt{|\alpha|})} + M) + 4 M (|\alpha| - m \sqrt{|\alpha|}) \exp(-\frac{m^2}{3})$ with $m = |\alpha|^{1/3}$. 

As an example consider a solenoid with radius $.7$ cm and length $4$ cm, the volume is $2 \pi \times 10^{-6}$,
$m^3$. Rabi frequency of $100$ kHz corresponds to field strength of $2.35 * 10^-3$ Tesla. The stored energy is $V \frac{B^2}{\mu_0}$. This corresponds to $n \hbar \omega$ photon energy units. This means $n = 10^{24}$. For $M < 10^{10}$, we have
$\frac{M^3}{\langle n \rangle ^2}$ and $\frac{M}{\langle n \rangle }$ are negligible. The total error is negligible. We do get a perfect $\frac{\pi}{2}$ pulse as governed by Bloch equations in magnetic resonance.

\section{Evolution of Boltzmann polarized spin ensemble in coherent rf-field}

We now consider a spin ensemble of $M_1 + M_2$ spins, $M_1$ polarized in up direction and $M_2$ polarized in down direction with $M_1 - M_2$ as the net polarization. We begin by interacting with a electromagnetic field in number state $| n \rangle$.
We choose a basis for the spin states as
$$ |k, j \rangle = \frac{1}{\sqrt{M_1 \choose {k+j}} \sqrt{M_2 \choose {k}}}| 0^{M_1-(k+j)} 1^{k + j} \rangle | 0^{k} 1^{M_2 - k} \rangle, $$

For $j \leq n$, the transition rate from state $|n-j-1 \rangle|k, j-1 \rangle$ and $|n-j+1 \rangle |k-1, j+1 \rangle$ to $|n-j \rangle |k, j \rangle $ is

$$\kappa \sqrt{n-j} \sqrt{(k+j) (M_1-k-j+1)} = \kappa \sqrt{n} (1 -\frac{j}{n}) \sqrt{(k+j) (M_1-k-j+1)} \sim \kappa \sqrt{n} \sqrt{(k+j) (M_1-k-j+1)}, $$ and

$$\kappa \sqrt{n-j +1} \sqrt{k (M_2-k+1)} = \kappa \sqrt{n}(1 - \frac{j-1}{n}) \sqrt{k (M_2-k+1)} \sim \kappa \sqrt{n} \sqrt{k (M_2-k+1)},$$
respectively. The transition rate from $|n-j \rangle |k, j \rangle$ to state $|n-j-1 \rangle|k, j+1 \rangle $ and $|n-j+1 \rangle |k+1, j-1 \rangle$ is

$$\kappa \sqrt{n-j} \sqrt{(k+j+1) (M_1-k-j)} = \kappa \sqrt{n}(1-\frac{j}{n}) \sqrt{(k+j+1) (M_1-k-j)} \sim \kappa \sqrt{n} \sqrt{(k+j+1) (M_1-k-j)}, $$ and

$$\kappa \sqrt{n+1-j} \sqrt{(k+1) (M_2-k)} = \kappa \sqrt{n} (1 - \frac{j-1}{n}) \sqrt{(k+1) (M_2-k)} \sim \kappa \sqrt{n} \sqrt{(k+1) (M_2-k)} $$ respectively

Let $a_{k,j}$ be the coefficient of the state $|k, j \rangle$. Then

\begin{eqnarray}
\frac{d a_{k,j}}{dt} &=& -i \omega (n + 1 - \frac{M_1-M_2}{2})a_{k,j} -i \kappa \sqrt{n} ( \sqrt{(k+j) (M_1-k-j+1)} a_{k, j-1} \\ \nonumber
&+& \sqrt{k (M_2-k+1)} a_{k-1, j+1} + \sqrt{(k+j+1) (M_1-k-j)} a_{k, j+1} + \sqrt{(k+1) (M_2-k)} a_{k+1, j-1}).
\end{eqnarray}

$b_{k,j} = \exp(-i (2*k + j)\frac{\pi}{2}) a_{k,j}$, Then writing equation for
$b_{k,j}$,

\begin{eqnarray}
\frac{d b_{k,j}}{dt} &=& -i \omega (n + 1 - \frac{M_1- M_2}{2})b_{k,j} + \kappa \sqrt{n} ( - \sqrt{(k+j) (M_1-k-j+1)} b_{k, j-1} \\ \nonumber &-& \sqrt{k (M_2-k+1)} b_{k-1, j+1} + \sqrt{(k+j+1) (M_1-k-j)} b_{k, j+1} + \sqrt{(k+1) (M_2-k)} b_{k+1, j-1}).
\end{eqnarray}

The solution to the above differential equation can be written as
$$ b_{k j}(\sigma) = \sqrt{M_1 \choose k + j} \sqrt{M_2 \choose k} \exp(-i \omega t (n + 1 - \frac{M_1- M_2}{2})) (-1)^j \cos^{M_1 + M_2 -(2k + j)}(\sigma(t)) \sin^{2k + j}(\sigma(t)). $$

$$ b_{k j}^2(\sigma) = {M_1 \choose k + j} {M_2 \choose k} (\underbrace{\cos^2 \sigma(t)}_p)^{M_1 + M_2 -(2k + j)}(\underbrace{\sin^2 \sigma(t)}_q )^{2k + j}.$$

where for $\sigma(t) = \kappa \sqrt n t$ the coefficient,

\begin{eqnarray*}
\frac{d b_{k j}}{dt} &=& -i \omega (n + 1 - \frac{M-1 - M_2}{2})b_{k,j} - \sqrt{(k+j)(M_1-(k+j)+1)} b_{k j-1} - \sqrt{k(M_2 -k +1)}b_{k-1, j+1} \\ &+& \sqrt{(M_1-(k+j))(k+j+1)} b_{k, j+1} + \sqrt{(k+1)(M_2-k)}b_{k+1, j-1}.
\end{eqnarray*}

The solution to the above differential equation can be described as follows. $\bar a_{k,j}(t)$ is the coefficient of $|k, j \rangle$ in,

$$ (\cos \kappa \sqrt n t |0 \rangle - i\sin \kappa \sqrt n t |1 \rangle )^{\otimes M1}(\cos \kappa \sqrt n t |1 \rangle - i \sin \kappa \sqrt n t |0 \rangle )^{\otimes M2}, $$ and $a_{k,j}(t) = \exp(-i \omega t (n + 1 - \frac{M_1-M_2}{2})\bar a_{k,j}(t)$.

If the initial state of the field is a coherent state, with parameter $\alpha = |\alpha| e^{-j \theta}$, then the state evolves as
\begin{equation}
\label{eq:nme}
 \sum_j (\sum_{n} \exp(-i \omega t (n + j + 1 - \frac{M_1-M_2}{2}) c_{n+j} |n \rangle) \sum_k \bar a_{k, j}(t)|k, j \rangle ).
 \end{equation}

Let

$$ |\psi_1 \rangle = \sum_j \exp(-i \omega t (j + 1 - \frac{M_1 - M_2}{2}) + j \theta ) (\sum_{n} \exp(-i \omega t (n+\theta) ) \frac{|\alpha|^{\frac{n+j}{2}}}{\sqrt{n+j!}} \exp(-\frac{|\alpha|}{2}) | n \rangle) \sum_k \bar a_{k, j}(t)|k, j \rangle. $$

For $n$ near mean photon number $|\alpha| = \langle n \rangle$, we have
$\frac{|\alpha|^{\frac{j}{2}}}{\sqrt{(n+j)\dots (n+1)}} \sim 1$,

Let 
$$ |\psi_2 \rangle = \sum_j \exp(-i \omega t (1 - \frac{M_1-M_2}{2})) (\sum_{n}  \exp(-i \omega t (n+\theta) ) \frac{|\alpha|^{\frac{n}{2}}}{\sqrt{n!}} \exp(-\frac{|\alpha|}{2}) | n \rangle )
\sum_k \bar a_{k, j}(t)| k, j \rangle \exp(-i (\omega t + \theta) j ). $$

As in Eq. \ref{eq:inpro},

$$ \langle \psi_1 | \psi_2 \rangle \geq (1 - \epsilon) \underbrace{\sum_{E = \{ |j| \leq |\alpha|^{\frac{1}{2}-\beta} \}}|a_{k,j}|^2}_{Pr(X_1-X_2) \leq |\alpha|^{\frac{1}{2}-\beta} }   + \sum_{E^c}, $$

where $X_1 = k+j$ and $X_2 = k$. Then we can treat $X_1$ and $X_2$ are independent Random variables with $|a_{k,j}|^2$ as there joint probability. $X_1-X_2$ has mean $(M_1-M_2)p$ and variance $(M_1 + M_2)pq$. At terminal time $t = \frac{\pi}{4 \kappa \sqrt{\alpha}}$,
$p, q = \frac{1}{2}$. For example, when $M_1 = 10^{15}$, $M_1 - M_2 = 10^{10}$, $\alpha = 10^{24}$, $\beta = \frac{1}{12}$, we have $Pr(E) > 1 - \epsilon'$, with $\epsilon'$ negligible.

The spin-photon state takes the form 
$$ |\psi(t)\rangle (\cos \kappa \sqrt n t |0 \rangle - i \exp(-i (\omega t + \theta) ) \sin \kappa \sqrt n t |1 \rangle )^{\otimes M_1} (\cos \kappa \sqrt n t |1 \rangle -i  \exp(i (\omega t + \theta) ) \sin \kappa \sqrt n t |0 \rangle )^{\otimes M_2}. $$

where $|\psi(t)\rangle$ is coherent state with parameter $\alpha(t) = |\alpha| \exp(i 2 (\omega t + \theta)).$ When evolution time is chosen such that $t = \frac{\pi}{4 \kappa \sqrt{\langle n \rangle}}$, the state takes the form

$$ |\psi(t)\rangle (\frac{|0 \rangle - i \exp(-i ( \omega t + \theta))|1 \rangle}{\sqrt 2} )^{\otimes M1} (\frac{|1 \rangle- i \exp(i ( \omega t + \theta))|0 \rangle}{\sqrt 2} )^{\otimes M2}. $$

The state of field and spin ensemble is a separable state.

\subsection{Error due to variation in coupling strength to different number states}

As before when deriving evolution of coherent field and spin ensemble we make the approximation that
$\sqrt{n} \sim \sqrt{\langle n\rangle}$. We can capture the error due to approximation in the evolution. Let $\Delta_{kj}^n = \tilde{b}_{k,j} - b_{k,j}$

\begin{eqnarray*}
\frac{d b_{k j}}{dt} &=& -i \omega (n + 1 - \frac{M_1-M_2}{2})b_{k,j} -(\kappa \sqrt{n})(\sqrt{(k+j)(M_1-(k+j)+1)} b_{k j-1} + \sqrt{k(M_2 -k +1)}b_{k-1, j+1} \\ &-& \sqrt{(M_1-(k+j))(k+j+1)} b_{k, j+1}^2 - \sqrt{(k+1)(M_2-k)}b_{k+1, j-1}) \\
|(\Delta b)^n_{kj}|^2 &=&  \kappa^2 ((\sqrt{n-j}-\sqrt n)^2 + (\sqrt{n}-\sqrt \alpha)^2) \{  (k+j)(M_1-(k+j)+1) b_{k, j-1}^2 + k(M_2 -k +1)b_{k-1, j+1}^2 \\ &+& (M_1-(k+j))(k+j+1) b_{k, j+1}^2 + (k+1)(M_2-k)b_{k+1, j-1} \\ &-& 2 \sqrt{(k+j)(M_1-(k+j)+1)(M_1-(k+j))(k+j+1)} b_{k,j}^2 \\ &-& 2 \sqrt{k(M_2 -k +1)(k+1)(M_2-k)}  b_{k,j}^2 \\ &+&
(( b_{k-1,j+1}^2 (k+j)(M_2-k+1) - b_{k, j}^2 (k+j)(M_2-k))) \\
&+& ( \sqrt{(k+j+1)(M_1-(k+j))} ( b_{k+1, j-1}^2 (k+1)(M_1-(k+j))- b_{k, j}^2 k(M_1-k+j))) \}.
\end{eqnarray*}

\begin{eqnarray*}
\frac{1}{T^2}\sum (\Delta_{kj}^n)^2 &=&  \frac{\kappa^2 n}{n^2}\{  \underbrace{j^2 (k+j)(M_1-(k+j)+1) b_{k, j-1}^2}_{E(X_1-X_2)^2(X_1+1)(M_1-X_1)} + \underbrace{k(M_2 -k +1)b_{k-1, j+1}^2}_{E(X_1-X_2)^2(X_2+1)(M_2-X_2)} \\ &+& \underbrace{(M_1-(k+j)(k+j+1) b_{k, j+1}^2}_{E(X_1-X_2)^2(X_1)(M_1-X_1 +1)} + \underbrace{(k+1)(M_2-k)b_{k+1, j-1}}_{E(X_1-X_2)^2(X_2)(M_1-X_2 +1)} \\ &-& 2 \underbrace{\sqrt{(k+j)(M_1-(k+j)+1)(M_1-(k+j))(k+j+1)} b_{k,j}^2}_{E(X_1-X_2)^2 X_1 (M_1 -X_1)} \\ &-& 2 \underbrace{\sqrt{k(M_2 -k +1)(k+1)(M_2-k)}  b_{k,j}^2}_{E(X_1-X_2)^2 X_2 (M_2 -X_2)} \\ &+&
\underbrace{b_{k-1,j+1}^2 k+j(M_2-k+1)}_{E((X_1-X_2 -1)^2 X_1(M_2 -X_2))} 
- \underbrace{(k+j)(M_2-k)b_{k, j}^2}_{E((X_1-X_2)^2 X_1(M_2 -X_2))} \\&+&
\underbrace{b_{k+1, j-1}^2 (k+1)(M_1-k -j)}_{E((X_1-X_2+1
)^2 (M_1-X_1)(X_2))} - \underbrace{b_{k, j}^2 k(M_1-k-j)}_{E((X_1-X_2)^2 (M_1-X_1)(X_2))} \}\\ \\ \\
&&\kappa^2 \alpha (\sqrt{n}-\sqrt \alpha)^2) 
\{\underbrace{(k+j)(M_1-(k+j)+1) b_{k, j-1}^2}_{E((X_1+1)(M_1-X_1))} + \underbrace{k(M_2 -k +1)b_{k-1, j+1}^2}_{E((X_2+1)(M_2-X_2))} \\ &+& \underbrace{(M_1-(k+j)(k+j+1)) b_{k, j+1}^2}_{E((X_1)(M_1-X_1 +1))} + \underbrace{(k+1)(M_2-k)b_{k+1, j-1}}_{E((X_2)(M_1-X_2 +1))} \\ &-& 2 \underbrace{\sqrt{(k+j)(M_1-(k+j)+1)(M_1-(k+j))(k+j+1)} b_{k,j}^2}_{E(X_1 (M_1 -X_1))} \\ &-& 2 \underbrace{\sqrt{k(M_2 -k +1)(k+1)(M_2-k)}  b_{k,j}^2}_{E(X_2 (M_2 -X_2))} \\ &+& \underbrace{b_{k-1,j+1}^2 k+j(M_2-k+1)}_{E(X_1(M_2 -X_2))} - \underbrace{(k+j)(M_2-k)b_{k, j}^2}_{E(X_1(M_2 -X_2))} \\&+&
\underbrace{b_{k+1, j-1}^2 (k+1)(M_1-k -j)}_{E((M_1-X_1)(X_2))} - \underbrace{b_{k,j}^2 k(M_1-k-j)}_{E((M_1-X_1)(X_2))} \}.
\end{eqnarray*}

The second part sums to $\kappa^2 (\sqrt{n}-\sqrt \alpha)^2(M_1 + M_2)$.

\begin{eqnarray*}
E(X_1-X_2)^2(X_1+1)(M_1-X_1) + E(X_1-X_2)^2(X_2+1)(M_2-X_2) &+& \\ E(X_1-X_2)^2(X_1)(M_1-X_1 +1) &+& \\
E(X_1-X_2)^2(X_2)(M_1-X_2 +1) -2 E(X_1-X_2)^2 X_1 (M_1 -X_1) -2 E(X_1-X_2)^2 X_2 (M_2 -X_2) &=& \\
2(M_1 + M_2) E(X_1-X_2)^2 &&.
\end{eqnarray*}

The total error has the pairing $\frac{(M_1-M_2)^2}{n}$ and $\frac{M_1}{n}$.

From Equation $\ref{eq:nme}$, the state evolves as

$$ \sum_j (\sum_{n} \exp(-i \omega t (n + j + 1 - \frac{M_1-M_2}{2}) c_{n+j} |n \rangle) \sum_k \bar a_{k, j}(t)|k, j \rangle) $$

We capture the error in the evolution as

\begin{eqnarray*}&&\sum_j (\sum_{n} \exp(-i \omega t (n + j + 1 - \frac{M_1 -M_2}{2}) c_{n+j} |n \rangle) \sum_k \bar a_{k, j}(t) |k, j \rangle) \\
&+& \underbrace{\sum_j (\sum_{n} c_{n+j} |n \rangle) \sum_k \exp(i \frac{\pi}{2} (2k+ j) \Delta^{n+j}_{k, 
j} |k, j \rangle)}_{|\phi \rangle}. \end{eqnarray*}

$$|\ |\phi \rangle|^2 = \sum_{n}\sum_{k, j}c_{n+j}^2 (\Delta^{n+j}_{k,j})^2. $$

Let $\epsilon_n^2 = \sum_{k,j} (\Delta^n_{k,j})^2$,

\begin{equation}
\label{eq:err2}
 |\  | \psi \rangle |^2 = \sum_n c_n^2 \epsilon_{n}^2  = \sum_{n \geq \alpha - m \sqrt{|\alpha|} }c_{n}^2 \epsilon_{n}^2 + \sum_{0}^{\alpha - m \sqrt{|\alpha|}} c_{n}^2 \epsilon_{n}^2.
\end{equation}

$$\epsilon_n^2 = o(\frac{(M_1-M_2)^2}{n}\frac{(M_1)}{n}). $$

The factors in Eq. \ref{eq:err2} are bounded by $o(\frac{(M_1-M_2)^2}{|\alpha|}\frac{(M_1)}{|\alpha|}) + 4 M_1 (|\alpha| - m \sqrt{|\alpha|}) \exp(-\frac{m^2}{3})$ with $m = |\alpha|^{\frac{1}{3}}$.
For example when $M_1 = 10^{15}$, $M_1 - M_2 = 10^{10}$, $|\alpha| = 10^{24}$, error is negligible.

\newpage

\section{Discussion}

We studied the evolution of a coherent rf-field with a 
sample of all polarized spins. We showed that evolution can be solved in closed form as a separable state
of rf-field and spin ensemble, where spin ensemble evolves according to Bloch equations in an rf field.
The rabi frequency and coupling strength of spins to rf-field
depends on number state of the rf-field. We showed that in interaction with a coherent rf-field, this variation in coupling strength, 
introduces negligible error.

\end{document}